\documentclass[a4paper,twocolumn,11pt]{quantumarticle}
\pdfoutput=1

\usepackage[english]{babel}
\usepackage{amsmath}
\usepackage{amssymb}
\usepackage{mathtools}
\usepackage{booktabs}
\usepackage{graphicx}
\usepackage{microtype}
\usepackage{hyperref}

\newcommand{\dd}{\mathrm{d}}
\newcommand{\e}{\mathrm{e}}
\newcommand{\ket}[1]{|#1\rangle}
\newcommand{\bra}[1]{\langle #1|}
\newcommand{\braket}[2]{\langle #1|#2\rangle}

\begin{document}

\title{Can a Measurement Be Undone? Recovering the State of a Measured Microscopic System with a Reversible Measuring Apparatus}

\author{Peter Renkel}
\affiliation{The MathWorks, Inc., Natick, MA, USA}

\date{July 26, 2026}

\begin{abstract}
We propose a direct, model-independent search for irreversible coherence loss
during a reversible measurement.
A mesoscopic apparatus measures a microscopic two-state system by storing
which-state information and is then returned to its pre-measurement state;
unitary quantum mechanics predicts recovered coherence, while collapse leaves
residual endpoint loss.

In one representative device, a coherently controlled
molecular force source displaces a charged nanoparticle that serves as the
apparatus; during one \(4.19\,{\rm ms}\) measurement-and-reversal cycle, its
two states become almost fully distinguishable.

Such a device would give a first direct bound in a measurement-and-reversal
setting. Repeating this cycle \(5\times10^5\) times would bound the
corresponding irreversible coherence-loss rate beyond ordinary decoherence at
\(5.51\,{\rm s}^{-1}\). Continuous Spontaneous Localization is included as a
secondary benchmark on the same apparatus history.
\end{abstract}

\maketitle

\section{Introduction}

The measurement problem is often stated as a contrast between two descriptions.
Before a measurement, quantum theory evolves a state unitarily. During a
measurement, a microscopic system becomes entangled with many degrees of
freedom of a measuring apparatus, and the corresponding apparatus states rapidly
become almost orthogonal \cite{Zurek1981,Schlosshauer2005}. The reduced density
matrix of the microscopic system then looks incoherent even in ordinary quantum
mechanics. This makes a direct test of collapse during measurement difficult:
loss of coherence in the microscopic system alone does not prove that a
physical collapse has occurred \cite{Schlosshauer2005,BassiRMP2013}.

The route proposed here is to make the measuring apparatus reversible.
Figure~\ref{fig:source_geometry} shows the representative setting used below: a
two-state polar molecule produces branch-dependent forces on a charged
nanoparticle, which serves as the measuring apparatus. A microscopic two-state
system pushes this mesoscopic apparatus in a branch-dependent way, so the
apparatus temporarily carries which-state information. Once that information
would be sufficient, in principle, to identify the microscopic alternative with
high probability, we regard the interaction as a measurement. Instead of reading
out the apparatus, the dynamics is closed by a harmonic phase-space loop so that
the apparatus returns to its initial state before readout.

If standard
quantum mechanics is the complete description, the microscopic interference
must return except for ordinary environmental decoherence and controlled
technical imperfections. If a collapse process occurs while the apparatus
carries which-state information, the recovery is incomplete.

Earlier quantum-eraser and measurement-reversal work shows how ordinary
quantum information can be erased or conditionally unwritten
\cite{Chapman1995,Ueda1996,KorotkovJordan2006,Katz2008}. Deutsch discussed
related thought experiments in the Everett (many-worlds) interpretation, where
recoherence after reversing a measurement follows from universal unitary
evolution \cite{Deutsch1986}. Here those ideas are turned into a mesoscopic
test of whether microscopic coherence returns after the apparatus information
is removed. The new ingredient is that the information is first stored in a
mesoscopic measuring apparatus and then coherently removed before readout. A
statistically significant residual loss of microscopic coherence would therefore be a direct
signal of physics beyond ordinary reversible entanglement.

This distinguishes the proposal from direct matter-wave interferometry. In a
matter-wave test, including the recent sodium-cluster benchmark of
Pedalino et al.~\cite{Pedalino2026}, one tests whether the spatial coherence of a
single delocalized object survives separation and recombination. Here the
microscopic system and the mesoscopic apparatus are different physical objects:
the nanoparticle temporarily stores information about the molecule's internal
state and is then restored before the molecule is read out. The value of the
experiment is that it probes a different physical channel: reversible formation
and removal of apparatus-held which-state information.

This endpoint-recovery test is broader than a bound on any particular collapse
model. It asks whether microscopic coherence returns after a mesoscopic
apparatus has stored and then removed which-state information. Within this
operational class, a residual loss would constrain a broad class of objective
mechanisms that produce irreversible dephasing during the reversible
measurement, including mechanisms not yet captured by a standard named model.

That breadth comes with a tradeoff. Once a specific model is assumed, one can
also use its other predicted signatures, such as heating, force noise, or
spontaneous radiation, and those indirect channels often give numerically
stronger exclusions
\cite{Vinante2016,Vinante2017,Carlesso2016,Vinante2020,Donadi2021,Majorana2022,XENONnT2026,Carlesso2022}.
The proposed experiment is therefore a direct test of measurement reversibility,
complementary to the strongest model-specific searches.

Named collapse models are still useful as benchmarks. CSL and related
objective-collapse models
\cite{GRW1986,Pearle1989,GhirardiPearleRimini1990,BassiGhirardi2003,BassiRMP2013}
map an apparatus history to a predicted excess coherence loss. This paper gives
an explicit CSL benchmark because CSL provides the standard reference scale for
direct coherence-loss searches. DP-type kernels \cite{Diosi1989,Penrose1996}
can be evaluated on the same trajectory once a short-distance mass-density
regularization is specified.

\subsection{Protocol overview}
\label{sec:overview}

In the representative geometry of Fig.~\ref{fig:source_geometry}, a Ramsey
preparation pulse first creates a coherent superposition of two stationary
internal states of a polar molecule \cite{Ramsey1950}. Because the two states have different
dipole moments, this is also a dipole-state superposition. The two
branches therefore exert slightly different forces on the charged nanoparticle.

\begin{figure}[!t]
\centering
\includegraphics[width=\linewidth]{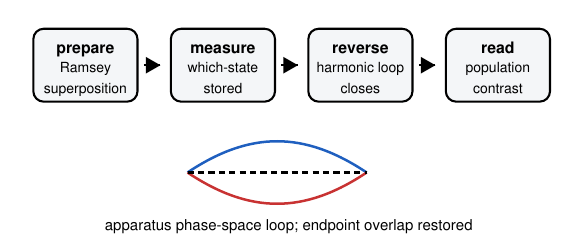}
\caption{Measurement-and-reversal cycle. The nanoparticle trajectories store
which-state information and then recombine before Ramsey readout.}
\label{fig:protocol}
\end{figure}

During the reversible loop, those branch-dependent forces separate the
nanoparticle trajectories in phase space. The nanoparticle motion then contains
which-state information, as summarized in Fig.~\ref{fig:protocol}. The loop is
chosen so that the two apparatus branches return to the same position and
momentum; the final nanoparticle displacement is designed to be zero.

A Ramsey-style final analysis pulse then rotates the two-state molecule so that
recovered microscopic coherence appears as different probabilities for
detecting the two internal molecular states. The operational question is
whether, after ordinary decoherence and control errors have been bounded, any
statistically significant residual loss remains. We parameterize such a
residual by the coherence factor \(\exp[-\Lambda_{\rm c}]\).

The estimates assume that a possible collapse contribution appears as extra
dephasing between the two internal molecular states used in the Ramsey sequence,
while ordinary decoherence and control loss are independently bounded and
accounted for as \(\Lambda_{\rm loss}\). We characterize one reversible loop by
how well the two apparatus states can be distinguished during the loop,
\(\mathcal{D}_{\rm exp}(t)\), and by its time integral,
\(J_{\rm exp}=\int_0^\tau \mathcal{D}_{\rm exp}(t)\,\dd t\), where \(\tau\) is
the loop duration. For the RF-compatible representative point,
\(\max_t\mathcal{D}_{\rm exp}\simeq1.00\) and
\(J_{\rm exp}\simeq0.832\tau\). With \(5\times10^5\) complete
measurement-and-reversal cycles, the smallest detectable extra loss rate per
unit \(J_{\rm exp}\) is \(\Gamma_{\rm exp}^{\min}=5.51\,{\rm s}^{-1}\).
Specific models enter by translating the same apparatus history into
\(\Lambda_{\rm c}\). In CSL, included as a familiar scale-setting example, the
same point reaches
\(\lambda_{\min}\simeq2.6\times10^{-11}\,{\rm s}^{-1}\) at
\(r_c=100\,{\rm nm}\), using the exact finite-width CSL expression in
Appendix~\ref{app:finitewidth}.

\section{Reversible Measurement Protocol}
\label{sec:protocol}

Let \(\ket{1}\) and \(\ket{2}\) denote the two internal molecular states used as
the measured system, and let \(\chi\) label this two-state molecular subsystem.
Its density matrix is \(\rho_\chi\).
After the first Ramsey pulse the molecule is prepared in the equal
superposition \((\ket{1}+\ket{2})/\sqrt2\), so
\begin{equation}
  \rho_\chi(0)=\frac{1}{2}
  \left(\ket{1}\bra{1}+\ket{2}\bra{2}
  +\ket{1}\bra{2}+\ket{2}\bra{1}\right).
\end{equation}
The nanoparticle apparatus has coordinate \(x\), momentum \(p\), and mass
\(m\). It is first equilibrated in a tight preparation trap of frequency
\(\omega_0\) at temperature \(T\), and then evolves during the measurement in a
weak trap of frequency \(\omega\). For a high-temperature thermal preparation,
the position width during the weak-trap interval is
\begin{equation}
  \sigma^2(t)=
  \frac{k_{\rm B}T}{m\omega_0^2}\cos^2\omega t+
  \frac{k_{\rm B}T}{m\omega^2}\sin^2\omega t .
  \label{eq:breathing}
\end{equation}
The full phase-space covariance is given in Appendix~\ref{app:thermalwidth}.

\begin{figure*}[!t]
\centering
\includegraphics[width=0.96\textwidth]{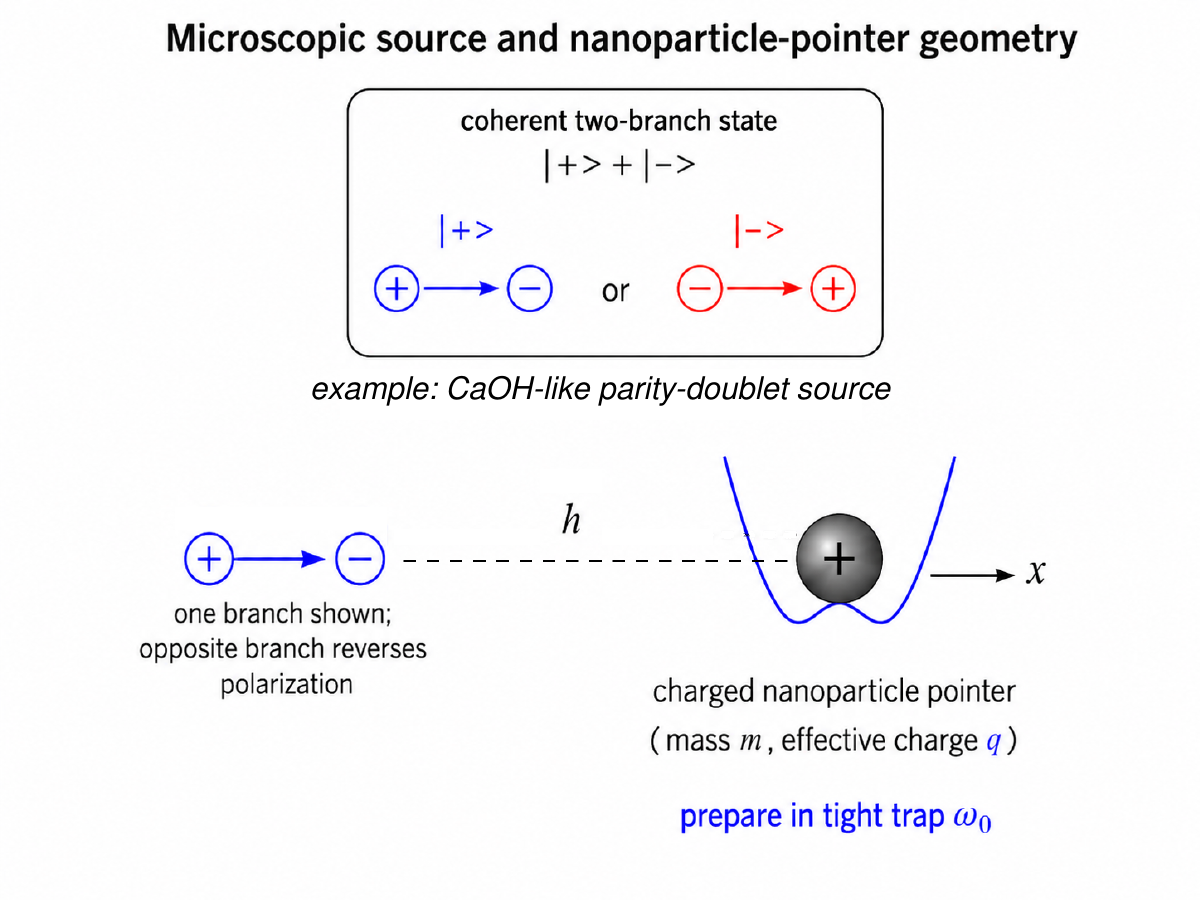}
\caption{Concrete source-apparatus geometry. A CaOH-like parity-doublet molecule
creates a known differential force \(\Delta F\) on the charged nanoparticle
apparatus; the conservative benchmark uses \(\Delta p=1.5\,{\rm D}\).}
\label{fig:source_geometry}
\end{figure*}

For the concrete estimates we use the source-apparatus geometry of
Fig.~\ref{fig:source_geometry}. During the interaction the two microscopic
states exert forces
\begin{equation}
  F_1=\bar F+\frac{\Delta F}{2},\qquad
  F_2=\bar F-\frac{\Delta F}{2}.
\end{equation}
The branch Hamiltonians are
\begin{equation}
  H_i=\frac{p^2}{2m}+\frac{1}{2}m\omega^2x^2-F_i x,\qquad i=1,2 .
  \label{eq:Hbranches}
\end{equation}
The common force \(\bar F\) moves both branches together and contributes only a
deterministic phase. The differential force \(\Delta F\) generates the branch
separation.

For zero initial position and momentum means, each branch follows
\begin{equation}
  \begin{aligned}
  \langle x_i(t)\rangle
  &=\frac{F_i}{m\omega^2}\left(1-\cos\omega t\right),\\
  \langle p_i(t)\rangle&=\frac{F_i}{\omega}\sin\omega t ,
  \end{aligned}
\end{equation}
The signed branch separation is
\begin{equation}
  D(t)\equiv\langle x_1(t)\rangle-\langle x_2(t)\rangle
  =\frac{\Delta F}{m\omega^2}\left(1-\cos\omega t\right).
  \label{eq:DofT}
\end{equation}
The corresponding relative momentum is
\begin{equation}
  \Pi(t)\equiv\langle p_1(t)\rangle-\langle p_2(t)\rangle
  =m\dot D(t)=\frac{\Delta F}{\omega}\sin\omega t .
  \label{eq:PiOfT}
\end{equation}
The maximum separation and interaction time are
\begin{equation}
  D_{\max}=\frac{2\Delta F}{m\omega^2},\qquad
  \tau=\frac{2\pi}{\omega}.
\end{equation}
At the endpoint,
\begin{equation}
  \begin{aligned}
  \langle x_i(\tau)\rangle&=0,\qquad
  \langle p_i(\tau)\rangle=0,\\
  D(\tau)&=0,\qquad \Pi(\tau)=0.
  \end{aligned}
  \label{eq:recombination}
\end{equation}
Thus each branch returns to the initial phase-space point.

\subsection{Endpoint density-matrix readout}

Starting from the initial molecular state prepared above, the apparatus loop
adds a relative molecular phase \(\phi\). Ordinary decoherence and bounded
control imperfections contribute \(\Lambda_{\rm loss}\), while possible
collapse contributes \(\Lambda_{\rm c}\). The readout tests collapse models that
reduce the ensemble-averaged off-diagonal coherence in this basis. We write
\(\Lambda_{\rm tot}=\Lambda_{\rm loss}+\Lambda_{\rm c}\), so the off-diagonal
element evolves as
\begin{equation}
  \rho_{12}(\tau)=\rho_{12}(0)
  \e^{i\phi}\e^{-\Lambda_{\rm tot}} .
  \label{eq:coherence_evolution}
\end{equation}
Thus, after the apparatus branches have recombined and before the Ramsey
analysis pulse, the ensemble-averaged molecular density matrix over many
nominally identical repetitions is
\begin{equation}
  \rho_\chi(\tau)=
  \frac{1}{2}
  \begin{pmatrix}
  1 & \e^{-\Lambda_{\rm tot}}\e^{i\phi} \\
  \e^{-\Lambda_{\rm tot}}\e^{-i\phi} & 1
  \end{pmatrix}.
  \label{eq:endpointmatrix}
\end{equation}

The Ramsey analysis pulse is represented by
\begin{equation}
  U_\theta=\frac{1}{\sqrt{2}}
  \begin{pmatrix}
  1 & \e^{i\theta}\\
  -\e^{-i\theta} & 1
  \end{pmatrix}.
  \label{eq:ramseyunitary}
\end{equation}
It changes the molecular state to
\begin{equation}
  \rho_{\chi}^{\rm out}(\theta)=
  U_\theta\rho_\chi(\tau)U_\theta^\dagger .
  \label{eq:rhoout}
\end{equation}
The diagonal elements of \(\rho_{\chi}^{\rm out}\) are the measured output
probabilities,
\begin{equation}
  \begin{aligned}
  P_1(\theta)
  &=\frac{1}{2}\left[
  1+\e^{-\Lambda_{\rm tot}}\cos(\phi-\theta)
  \right],\\
  P_2(\theta)&=1-P_1(\theta).
  \end{aligned}
  \label{eq:ramseypop}
\end{equation}

By scanning the analysis phase \(\theta\), the experiment determines
\(\Lambda_{\rm tot}\). The independently bounded value of
\(\Lambda_{\rm loss}\) is then subtracted from the total exponent to obtain
\(\Lambda_{\rm c}\).

\section{Apparatus-Exposure Collapse Benchmark}
\label{sec:exposure}

Section~\ref{sec:protocol} defines the measured residual exponent
\(\Lambda_{\rm c}\). This number is not by itself a universal parameter: it
depends on how much which-state information the apparatus carried and for how
long. To compare reversible-measurement experiments, we define an apparatus
exposure: the time integral of one minus the squared overlap of the two
apparatus states.

During the loop, the two microscopic alternatives condition two possible
apparatus states, denoted by \(\ket{A_1(t)}\) and \(\ket{A_2(t)}\). We define
the apparatus exposure
\begin{equation}
  J_{\rm exp}
  =\int_0^\tau \mathcal{D}_{\rm exp}(t)\,\dd t,
  \label{eq:exposuredistinguishability}
\end{equation}
where
\[
  \mathcal{D}_{\rm exp}(t)
  =1-\left|\braket{A_1(t)}{A_2(t)}\right|^2 .
\]
We use \(\mathcal{D}_{\rm exp}\) as a bounded which-way exposure measure: it is
zero when the apparatus states are identical and one when they are orthogonal.
For the representative point in Table~\ref{tab:points}, the maximum value
rounds to 1.00, so an ideal readout would identify the
microscopic alternative with essentially unit probability at maximum separation
at the quoted precision.

For the Gaussian apparatus used in the estimates below, we use the
equal-covariance Gaussian extension
\begin{equation}
  \begin{aligned}
  \mathcal{D}_{\rm exp}(t)
  &=1-\exp[-Q_{\rm exp}(t)],\\
  Q_{\rm exp}(t)
  &=\frac{1}{4}\delta z^{\mathsf T}(t)V^{-1}(t)\delta z(t),\\
  \delta z(t)&=(D(t),\Pi(t))^{\mathsf T}.
  \end{aligned}
  \label{eq:thermalexposure}
\end{equation}
The covariance matrix in Eq.~\eqref{eq:thermalexposure} is the thermal
phase-space covariance of the nanoparticle after the trap switch; its components
are listed in Appendix~\ref{app:thermalwidth}.

With this exposure defined, we use a phenomenological collapse benchmark in
which the irreversible exponent is proportional to the exposure
\cite{BassiGhirardi2003,BassiRMP2013}:
\begin{equation}
  \begin{aligned}
  \rho_{12}(\tau)&=\rho_{12}^{\rm QM}(\tau)
  \exp[-\Gamma_{\rm exp}J_{\rm exp}],\\
  \Lambda_{\rm c}&=\Gamma_{\rm exp}J_{\rm exp}.
  \end{aligned}
  \label{eq:exposureexponent}
\end{equation}
Here \(\Gamma_{\rm exp}=\Lambda_{\rm c}/J_{\rm exp}\) is the residual collapse
exponent normalized by the apparatus exposure.
Standard quantum mechanics corresponds to \(\Gamma_{\rm exp}=0\): if the loop
is closed, the which-state information is erased and the final coherence
returns. A positive \(\Gamma_{\rm exp}\) gives residual dephasing even after the
apparatus is restored.

For the endpoint probability in Eq.~\eqref{eq:ramseypop}, the small-exponent
\(5\sigma\) reach derived in Appendix~\ref{app:shotnoise} is
\begin{equation}
  \Gamma_{\rm exp}^{\min}\simeq
  \frac{5\e^{\Lambda_{\rm loss}}}{\sqrt{N}\,J_{\rm exp}} .
  \label{eq:gammaexpmin}
\end{equation}
If the apparatus branches are almost fully distinguishable in this squared sense
for one
\(\tau=4.19\,{\rm ms}\) loop, then \(J_{\rm exp}\simeq\tau\) and
\(\Gamma_{\rm exp}^{\min}=4.59\,{\rm s}^{-1}\) for
\(N=5\times10^5\) and \(\Lambda_{\rm loss}=1.0\). This is a
complete-information reference value. For the representative benchmark in
Table~\ref{tab:points}, Eq.~\eqref{eq:thermalexposure} gives
\(\max_t\mathcal{D}_{\rm exp}\simeq1.00\) and
\[
  J_{\rm exp}\simeq3.49\times10^{-3}\,{\rm s}
  =0.832\tau .
\]
The corresponding projected reach is therefore
\(\Gamma_{\rm exp}^{\min}=5.51\,{\rm s}^{-1}\). The number of measurement
cycles only enters this statistical conversion; the exposure \(J_{\rm exp}\)
itself is a property of one reversible measurement cycle.
Equations~\eqref{eq:exposuredistinguishability}--\eqref{eq:gammaexpmin}
therefore translate the apparatus-specific exponent \(\Lambda_{\rm c}\) into a
bound on the normalized collapse exponent \(\Gamma_{\rm exp}\), which can be
compared across experiments using the same exposure convention.

Future variants of this benchmark could change the exposure functional itself.
Other collapse ideas could weight the participating mass, copied degrees of
freedom, the phase-space distance between the two nanoparticle apparatus states,
distinguishability thresholds, or hidden apparatus degrees of freedom not
reversed by the protocol. The observable remains the fitted endpoint exponent
\(\Lambda_{\rm c}\), divided by the exposure integral appropriate to that model.

\section{Model Scope}
\label{sec:beyond}

The apparatus-exposure rate \(\Gamma_{\rm exp}\) is the model-independent output
of the proposal within the convention of Eq.~\eqref{eq:exposureexponent}. The
experiment first fits the residual endpoint exponent \(\Lambda_{\rm c}\), and
then reports \(\Gamma_{\rm exp}=\Lambda_{\rm c}/J_{\rm exp}\), because
\(\Lambda_{\rm c}\) alone depends on the apparatus history. This rate is meant
for mechanisms in which irreversible coherence loss is tied to the
distinguishability history of the measuring apparatus.

More generally, any mechanism that irreversibly suppresses coherence while the
apparatus carries which-state information contributes to the endpoint exponent,
\begin{equation}
  \Lambda_{\rm c}=\int_0^\tau
  \mathcal{K}_{\rm model}[A_1(t),A_2(t)]\,\dd t,
\end{equation}
where \(\mathcal{K}_{\rm model}\) is the model-dependent kernel. The same
trajectory \(D(t)\), the same decoherence-and-control budget, and the same
final Ramsey readout can therefore be used to constrain other exposure-based
collapse models. The next section gives CSL as a familiar worked benchmark.
Other models can be considered in future work; DP-type benchmarks are a natural
example once a short-distance mass-density regularization is chosen.

\section{Illustrative CSL Benchmark}
\label{sec:csl}

CSL is included as a familiar scale-setting example because its established
dynamics give a coherence-loss exponent for the branch separation \(D(t)\).
For a pointlike mass distribution, the standard CSL master equation gives the
branch-coherence decay rate
\cite{GhirardiPearleRimini1990,BassiRMP2013,Nimmrichter2011}
\begin{equation}
  \Gamma_{\rm CSL}^{\rm point}(D)
  =\lambda\frac{m^2}{m_0^2}
  \left[1-\exp\!\left(-\frac{D^2}{4r_c^2}\right)\right],
  \label{eq:cslpoint}
\end{equation}
with nucleon mass \(m_0\). For the representative point in
Table~\ref{tab:points}, \(D_{\max}=6.4\,{\rm nm}\), while the force-enhanced
outlook has \(D_{\max}=64\,{\rm nm}\). For the finite prepared width
\(\sigma(t)\) in Eq.~\eqref{eq:breathing}, we use a conservative isotropic
three-dimensional Gaussian mass density and the exact finite-width expression
derived in Appendix~\ref{app:finitewidth}:
\begin{align}
  \Gamma_{\rm CSL}^{\rm 3D}(t)
  &=
  \lambda\frac{m^2}{m_0^2}
  \left(1+\frac{\sigma^2(t)}{r_c^2}\right)^{-3/2}
  \nonumber\\
  &\quad\times
  \left\{
  1-\exp\!\left[
  -\frac{D^2(t)}{4(r_c^2+\sigma^2(t))}
  \right]\right\}.
  \label{eq:cslfinitewidth}
\end{align}
We then define
\begin{equation}
  \begin{aligned}
  K_{\rm CSL}(r_c)
  &=
  \int_0^\tau
  \frac{\Gamma_{\rm CSL}^{\rm 3D}(t)}{\lambda}\,\dd t,
  \\
  \Lambda_{\rm CSL}(r_c)
  &=
  \lambda K_{\rm CSL}(r_c).
  \end{aligned}
  \label{eq:Kcsl}
\end{equation}
In this paper, CSL simply converts the already specified apparatus history into
a predicted coherence-loss exponent through Eq.~\eqref{eq:cslfinitewidth}.
Figure~\ref{fig:csl_comparison} places the resulting CSL benchmark alongside
direct interferometric and spontaneous-radiation bounds.

\section{Representative Sources and Reach}
\label{sec:source}

The concrete source model is a single trapped CaOH-like parity-doublet polar
molecule. Recent experiments prepared optically trapped CaOH molecules in
\(\ell\)-type parity-doublet states and measured a bare coherence time
\(T_2^*=0.8(2)\,{\rm s}\), far longer than the millisecond loop considered here
\cite{Robichaud2026}. Earlier work demonstrated quantum control of trapped CaOH
in modest polarizing fields and coherent internal-state superpositions
\cite{Anderegg2023}. We therefore assume the source molecule is held in an
optical tweezer while its internal dipole state is coherently controlled. The
force estimate below uses \(\Delta p=1.5\,{\rm D}\), a conservative order-D
molecular scale rather than an optimized value.

This source choice should be read operationally. The proposal requires a
coherent two-state source that produces a known differential force on the charged
nanoparticle. In a molecular implementation, polarizing or dressing fields are
needed to create the lab-frame dipoles. The charged nanoparticle must be shielded
or compensated from the common part of those fields, and any residual field,
gradient, or drift at the nanoparticle must be included in the control budget.
The full molecule--nanoparticle device therefore remains an implementation
target.

The same protocol can be implemented with any coherent two-state force source.
A charged particle shared between two electrostatic wells is one long-term
alternative: for \(d\ll h\),
\(\Delta F\simeq2k_e q q_s d/h^3\). Electron-on-helium and double-well gate
architectures illustrate this route \cite{Koolstra2026,Leinonen2026}, whose
central requirement is source-charge coherence during the force interval. The
benchmark below uses the molecular route and should be read as a representative
force-source benchmark rather than a complete electrode design.

\subsection{Benchmark parameters}

The protocol itself is defined in Sec.~\ref{sec:protocol}; here we specify the
parameters used for the numerical reach in Table~\ref{tab:points}. The
ordinary-loss budget and control handles are summarized in
Sec.~\ref{sec:systematics}.

For a charged nanoparticle with effective charge \(q\) at distance \(h\) on the
dipole axis, the differential force scale is
\begin{equation}
  \Delta F\simeq
  \frac{q\,\Delta p}{2\pi\epsilon_0h^3}.
  \label{eq:dipoleforce}
\end{equation}

The CaOH control fields are of order \(10\)--\(10^2\,{\rm V/cm}\) at the
molecule. These fields set the molecular dipole state; the residual field at
the nanoparticle is treated separately in the control budget of
Sec.~\ref{sec:systematics}.

We use
\begin{align}
  m&=2.0\times10^{-19}\,{\rm kg},&
  q&=100e,&
  h&=10\,\mu{\rm m}, \nonumber\\
  \omega&=1.5\times10^3\,{\rm s}^{-1}.
\end{align}
With \(\Delta p=1.5\,{\rm D}\), Eq.~\eqref{eq:dipoleforce} gives
\(\Delta F=1.44\times10^{-21}\,{\rm N}\). The initial preparation uses
\(\omega_0=10^7\,{\rm s}^{-1}\) and \(T=0.1\,{\rm K}\), the interaction time is
\(\tau=4.19\,{\rm ms}\), and the projected reaches use
\(N_0=5\times10^5\) and \(\Lambda_0=1.0\).
The integrated nanoparticle implementation is demanding rather than routine.
The charged-nanoparticle loop would require an optimized trap and
state-preparation protocol, while an RF trap implementation with \(q=100e\)
would need a separately optimized design.
Besides the representative operating point, Table~\ref{tab:points} includes a
force-enhanced outlook to show the scaling if source or geometry optimization
provided a tenfold larger calibrated differential force at the same charge
scale.
For the illustrative CSL benchmark, a shot-noise-limited Ramsey fringe-amplitude
measurement gives, for \(\Lambda_{\rm c}\ll1\), the \(5\sigma\) reach:
\begin{equation}
  \lambda_{\min}(r_c)\simeq
  \frac{5\e^{\Lambda_{\rm loss}}}{\sqrt{N}\,K_{\rm CSL}(r_c)}.
  \label{eq:lambdamin}
\end{equation}
Changing exposure or ordinary loss rescales this result as
\begin{equation}
  \lambda_{\min}(N,\Lambda_{\rm loss})\simeq
  \lambda_{\min}(N_0,\Lambda_0)
  \e^{\Lambda_{\rm loss}-\Lambda_0}
  \left(\frac{N_0}{N}\right)^{1/2}.
  \label{eq:rescale}
\end{equation}
Here \(N_0=5\times10^5\) and \(\Lambda_0=1.0\) are the reference shot number
and ordinary-loss budget used in Table~\ref{tab:points}.

\begin{table*}[!t]
\centering
\caption{Representative operating points at \(r_c=100\,{\rm nm}\). Both rows use
\(m=2.0\times10^{-19}\,{\rm kg}\), \(T=0.1\,{\rm K}\),
\(\omega=1.5\times10^3\,{\rm s}^{-1}\), \(N_0=5\times10^5\), and
\(q=100e\). The force-enhanced outlook assumes a tenfold larger calibrated
differential force at the same ordinary-loss budget, \(\Lambda_0=1.0\).}
\label{tab:points}
{\scriptsize
\setlength{\tabcolsep}{3pt}
\begin{tabular}{lcccccc}
\toprule
Point & \(\Delta F\) & \(D_{\max}\) & \(\max\mathcal{D}_{\rm exp}\) &
\(J_{\rm exp}/\tau\) & \(\Gamma_{\rm exp}^{\min}\) &
\(\lambda_{\min}\) \\
 & \({\rm N}\) & \({\rm nm}\) & & & \({\rm s}^{-1}\) &
\({\rm s}^{-1}\) \\
\midrule
Representative RF-compatible point & \(1.44\times10^{-21}\) & 6.4 & 1.00 &
0.832 & 5.51 & \(2.6\times10^{-11}\) \\
Force-enhanced outlook & \(1.44\times10^{-20}\) & 64 & 1.00 &
0.948 & 4.84 & \(2.7\times10^{-13}\) \\
\bottomrule
\end{tabular}
}
\end{table*}

\begin{figure*}[!t]
\centering
\includegraphics[width=0.96\textwidth]{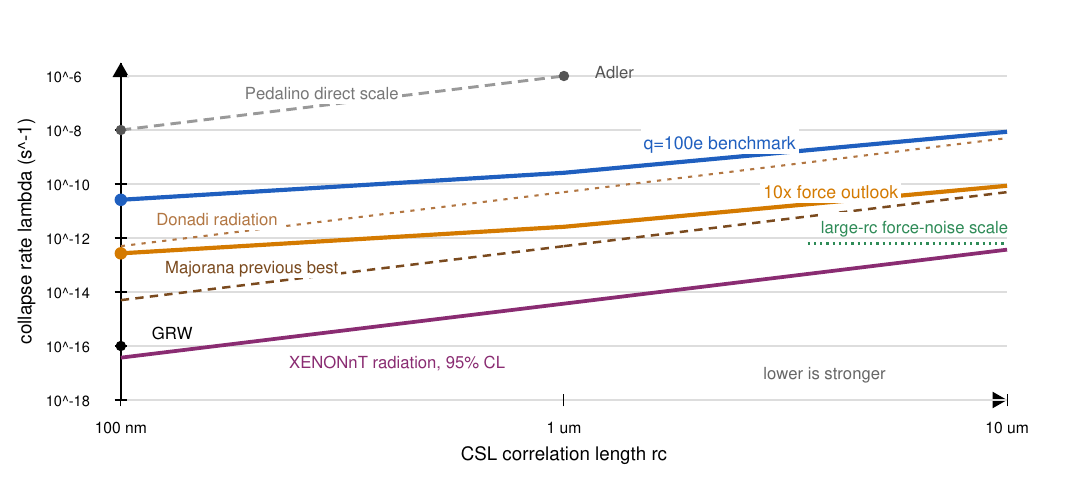}
\caption{Illustrative CSL comparison. Blue and orange show the \(q=100e\)
representative benchmark and force-enhanced outlook. Gray
marks the direct matter-wave scale from Pedalino et al.; tan, brown, and purple
show radiation benchmarks from Donadi et al., Majorana, and XENONnT.}
\label{fig:csl_comparison}
\end{figure*}

Table~\ref{tab:points} summarizes the result. The representative row
demonstrates the core reversible-measurement regime at a more realistic
nanoparticle charge scale: during one reversible loop the apparatus which-way
exposure measure rounds to \(1.00\), and the integrated exposure is
\(J_{\rm exp}\simeq0.832\tau\). These two numbers are properties of a single
cycle, fixed before choosing the shot number. The projected
\(\Gamma_{\rm exp}^{\min}\) and \(\lambda_{\min}\) columns use
\(N_0=5\times10^5\) to convert that physical exposure into a statistical reach
under the stated ordinary-loss convention. The residual exponent
\(\Lambda_{\rm c}\) and \(J_{\rm exp}\) are the central quantitative outputs.

CSL provides a familiar scale-setting use of the same branch history. Other
kernels, including a regularization-dependent DP benchmark, can be evaluated on
the same trajectory but are not part of the headline result. The force-enhanced
row shows how the reach would scale if source or geometry optimization provided
a tenfold larger coherent differential force while still requiring the
apparatus branches to recombine at the end.

Figure~\ref{fig:csl_comparison} presents the CSL comparison graphically. The
blue and orange curves are the \(q=100e\) representative benchmark and
force-enhanced outlook. The gray line is the direct matter-wave scale from
Pedalino et al.; the tan, brown, and purple curves are model-specific radiation
bounds from Donadi et al., Majorana, and XENONnT.

Pedalino et al. provide the direct massive-interference benchmark used in the
comparison, using sodium clusters with \(m\simeq170\,{\rm kDa}\), a
\(133\,{\rm nm}\) path separation, and macroscopicity \(\mu=15.5\)
\cite{Pedalino2026}. At \(r_c=100\,{\rm nm}\), the representative
reversible-measurement point reaches
\(\lambda_{\min}=2.6\times10^{-11}\,{\rm s}^{-1}\), more than two orders of
magnitude below the Pedalino direct scale shown in
Fig.~\ref{fig:csl_comparison}; the force-enhanced outlook reaches
\(\lambda_{\min}=2.7\times10^{-13}\,{\rm s}^{-1}\).

In the radiation channel, Donadi et al. found
\(\lambda<5.2\times10^{-13}\,{\rm s}^{-1}\) at \(r_c=100\,{\rm nm}\)
\cite{Donadi2021}; Majorana later reached
\(\lambda/r_c^2\simeq4.9\times10^{-1}\,{\rm s}^{-1}{\rm m}^{-2}\)
\cite{Majorana2022}; and XENONnT gives
\(\lambda/r_c^2<3.7\times10^{-3}\,{\rm s}^{-1}{\rm m}^{-2}\) at \(95\%\)
C.L. \cite{XENONnT2026}. The important distinction is the physical channel:
radiation searches probe a specific white-CSL emission mechanism, while the
present proposal probes endpoint coherence recovery after a reversible
measurement.

\section{Ordinary-Loss Budget and Controls}
\label{sec:systematics}

The endpoint Ramsey scan gives a total loss exponent
\[
  \Lambda_{\rm tot}=\Lambda_{\rm loss}+\Lambda_{\rm c}.
\]
Calibration measurements first establish the ordinary-loss reference
\(\Lambda_{\rm loss}\), which includes environmental decoherence and controlled
imperfections. The remaining statistically significant excess is attributed to
collapse and denoted \(\Lambda_{\rm c}\).

The controls are organized by purpose. Runs with no apparatus branch separation
set the baseline final fringe amplitude; they capture gas scattering, blackbody
emission, source coherence loss, and trap noise. The closest baseline is a
zero-differential-force run, which keeps the nearby source and trap environment
as similar as possible while removing the state-dependent force.

Branch-specific effects are checked by switching the source polarization. This
exchanges \(F_1\) and \(F_2\) while keeping the geometry fixed, so ordinary
coherent phases reverse sign whereas a true loss of fringe amplitude does not.

Endpoint closure is checked separately by scanning the recombination time around
the nominal closure point. Any Ramsey-fringe amplitude loss from imperfect
endpoint closure is included in \(\Lambda_{\rm loss}\).

The collapse-search run uses the independently measured recombination time and
then measures the endpoint probabilities for the two molecular states. Any
fitted loss beyond the independently measured ordinary-loss reference defines
\(\Lambda_{\rm c}\).

The most demanding entry is electrostatic stability at the charged
nanoparticle. A static electric field \(E\) at the nanoparticle shifts the
weak-trap equilibrium by
\[
  x_E=\frac{qE}{m\omega^2}.
\]
For the benchmark parameters, an uncompensated \(10\,{\rm V/cm}\) field at the
nanoparticle would give \(x_E\simeq3.6\,{\rm cm}\), while a residual field
\(E_{\rm res}=2.8\times10^{-5}\,{\rm V/m}\) corresponds to
\(x_E=1\,{\rm nm}\). The common electric field at the nanoparticle must
therefore be cancelled or shielded to the residual level used in the chosen
operating point. Residual drift and branch-correlated fields are therefore key
ordinary-loss channels for the collapse search. For the \(q=100e\) benchmark,
electrostatics dominate the working \(\Lambda_{\rm loss}=1.0\)
ordinary-loss budget. This is why source-field leakage, field drift, and
relative-position noise are explicit budget items.

Table~\ref{tab:systematics} summarizes the main channels and representative
targets. Gas scattering, source dephasing, field noise, and endpoint mismatch
contribute to the ordinary endpoint-loss exponent \(\Lambda_{\rm loss}\);
source-distance stability limits the calibrated force exposure used to quote
\(J_{\rm exp}\) and \(K_{\rm CSL}\). The 0.05--0.1-level targets are
therefore design allocations chosen to keep each non-electrostatic effect
subdominant to the working \(\Lambda_{\rm loss}=1.0\) benchmark.

\begin{table*}[!t]
\centering
\caption{Representative decoherence, control-loss, and calibration targets.
Device-specific designs should replace these benchmark allocations by measured
spectra.}
\label{tab:systematics}
\small
\begin{tabular}{p{0.18\textwidth}p{0.42\textwidth}p{0.30\textwidth}}
\toprule
Channel & Main role & Representative target \\
\midrule
Gas and blackbody & baseline environmental decoherence during the loop &
\(\Lambda_{\rm gas}\simeq4.8\times10^{-2}\); blackbody loss negligible for
cryogenic/shielded operation \\
Residual electric field & static equilibrium shift from uncompensated
polarizing-field leakage; shot-to-shot field drift gives phase averaging &
\(x_E=1\,{\rm nm}\) corresponds to
\(E_{\rm res}=2.8\times10^{-5}\,{\rm V/m}\);
an uncompensated \(10\,{\rm V/cm}\) field gives \(x_E\simeq3.6\,{\rm cm}\);
dominant contribution to \(\Lambda_{\rm loss}\) \\
Finite-frequency electric field & intrashot electric-field noise, giving phase
noise and possible endpoint mismatch &
measured \(S_E(\Omega)\) must be integrated through the loop response and
included in the same electrostatic-loss calibration \\
Source-distance noise & uncertainty in the inferred differential force
\(\Delta F\propto h^{-3}\), and hence in \(J_{\rm exp}\) and \(K_{\rm CSL}\) &
allocating \(\lesssim0.05\) fractional error to the force-squared exposure implies
\(6\sigma_h/h\lesssim0.05\), hence \(\sigma_h\lesssim80\,{\rm nm}\) for
\(h=10\,\mu{\rm m}\);
one-sided \(\sqrt{S_h}\lesssim21\,{\rm nm}/\sqrt{\rm Hz}\) \\
Timing and trap frequency & failure of the two nanoparticle branches to
recombine exactly at the endpoint &
allocating \(\Lambda_\omega\lesssim0.1\), with
\(\Lambda_\omega\simeq8.5\epsilon_\omega^2\) and
\(\epsilon_\omega\simeq\delta t/\tau\), gives
\(\delta t<4.5\times10^{-4}\,{\rm s}\) \\
Source coherence & baseline molecular Ramsey fringe amplitude &
allocating \(\Lambda_{\rm src}\lesssim0.05\), with
\(\Lambda_{\rm src}\simeq\tau/T_2\), gives
\(T_2\gtrsim84\,{\rm ms}\) \\
\bottomrule
\end{tabular}
\end{table*}

\section{Discussion}

The proposed experiment tests a specific question: if a microscopic system has
become entangled with a mesoscopic measuring apparatus, does restoring the
apparatus also restore the microscopic interference? Standard quantum mechanics
answers yes, provided ordinary decoherence and endpoint mismatch are controlled.
A collapse mechanism tied to apparatus exposure predicts residual coherence
loss because which-state information physically existed in the apparatus, even
though the apparatus was later returned to its initial state.

The representative benchmark uses a CaOH-like parity-doublet source with
\(\Delta p=1.5\,{\rm D}\), a \(q=100e\) charged nanoparticle of mass
\(2.0\times10^{-19}\,{\rm kg}\), a \(10\,\mu{\rm m}\) source-apparatus distance,
\(T=0.1\,{\rm K}\), and a \(4.19\,{\rm ms}\) weak-trap loop. With the stated
thermal preparation, it gives a
high-distinguishability reversible measurement:
\(\max_t\mathcal{D}_{\rm exp}\simeq1.00\) for the one-minus-squared-overlap
measure and \(J_{\rm exp}\simeq0.832\tau\). A
\(5\times10^5\)-cycle projection with \(\Lambda_{\rm loss}=1.0\) gives
\(\Gamma_{\rm exp}^{\min}=5.51\,{\rm s}^{-1}\). Achieving this regime requires
a coherent parity-doublet force source, electrostatic compensation and
calibration at the charged nanoparticle, a charged nanoparticle loop, and
verified endpoint recombination in the same device. Such a device would make
the first direct measurement or bound of the apparatus-exposure collapse rate
\(\Gamma_{\rm exp}\).

The same branch history gives a direct CSL benchmark. Other kernels can be
evaluated on the same trajectory. However, force-noise searches and
CSL-induced spontaneous-radiation searches constrain model-specific secondary
effects, whereas the endpoint-recovery test asks whether microscopic coherence
returns after the measuring apparatus first stores and then erases which-state
information.

Future improvements are straightforward at the scaling level:
\(\Lambda_{\rm CSL}^{(0)}\propto\Delta F^2\omega^{-5}\), and \(J_{\rm exp}\)
grows as the apparatus branches become more distinguishable and longer lived,
provided endpoint systematics remain controlled.

\section*{Acknowledgments}

P.R. thanks Matteo Carlesso, Sandro Donadi, Marko Toros, and Nicholas
R. Hutzler for helpful comments and correspondence. P.R. is
especially grateful to Nicholas R. Hutzler for pointing out recent CaOH
parity-doublet results relevant to the molecular-source implementation, and to
Sandro Donadi for guidance on current CSL limits and the CSL benchmark used for
comparison. P.R. also thanks David C. Moore and Vadim Asnin for helpful
discussions.

\section*{Author Contributions}

P.R. conceived the protocol, selected the physical assumptions, and is
responsible for the calculations and claims. A large language model was used
for editorial condensation, LaTeX reformatting, and draft organization; the
author reviewed and remains responsible for the manuscript.

\section*{Data and Code Availability}

No new experimental data were generated for this proposal. The numerical
estimates in Table~\ref{tab:points} follow directly from the formulas and
parameter values stated in the text.

\appendix

\section{Compact Derivations}
\label{app:derivations}

\subsection{Thermal width after trap switch}
\label{app:thermalwidth}

The nanoparticle is equilibrated in the tight trap and then evolved in the weak
trap. In the high-temperature limit
\(\langle x^2(0)\rangle=k_{\rm B}T/(m\omega_0^2)\),
\(\langle p^2(0)\rangle=mk_{\rm B}T\), and
\(\langle xp+px\rangle=0\). The last equality follows because the thermal state
of the tight trap is time-reversal invariant, while \(xp+px\) is odd under
\(p\to-p\). The weak-trap Heisenberg evolution is
\begin{align*}
  x(t)&=x(0)\cos\omega t+\frac{p(0)}{m\omega}\sin\omega t,\\
  p(t)&=p(0)\cos\omega t-m\omega x(0)\sin\omega t .
\end{align*}
Substitution gives
\[
  \sigma^2(t)=
  \frac{k_{\rm B}T}{m\omega_0^2}\cos^2\omega t+
  \frac{k_{\rm B}T}{m\omega^2}\sin^2\omega t ,
\]
which is Eq.~\eqref{eq:breathing}.
For Eq.~\eqref{eq:thermalexposure}, the phase-space covariance matrix
\[
  V(t)=
  \begin{pmatrix}
  V_{xx}(t) & V_{xp}(t)\\
  V_{xp}(t) & V_{pp}(t)
  \end{pmatrix}
\]
has \(V_{xx}(t)=\sigma^2(t)\) and
\[
  \begin{aligned}
  V_{pp}(t)&=mk_{\rm B}T\cos^2\omega t
  +mk_{\rm B}T\frac{\omega^2}{\omega_0^2}\sin^2\omega t,\\
  V_{xp}(t)&=k_{\rm B}T\sin\omega t\cos\omega t
  \left(\frac{1}{\omega}-\frac{\omega}{\omega_0^2}\right).
  \end{aligned}
\]

\subsection{Finite-width CSL factor}
\label{app:finitewidth}

For two mass-density profiles \(\mu_1(\mathbf r)\) and \(\mu_2(\mathbf r)\),
the CSL rate is the squared distance between the two Gaussian-smeared mass
densities:
\begin{align}
  \Gamma_{\rm CSL}
  &=
  \frac{\lambda}{2m_0^2}
  \left(\mathcal I_{11}+\mathcal I_{22}-2\mathcal I_{12}\right),\nonumber\\
  \mathcal I_{ij}
  &=
  \int\dd^3r\,\dd^3r'\,
  \mu_i(\mathbf r)\mu_j(\mathbf r')
  \exp\!\left[-\frac{|\mathbf r-\mathbf r'|^2}{4r_c^2}\right].
  \label{eq:appIij}
\end{align}
This follows by writing
\(\Gamma_{\rm CSL}=\lambda(2m_0^2)^{-1}\int\dd^3s
(\bar\mu_1-\bar\mu_2)^2\) and choosing the smearing normalization
\(\int\dd^3s\,g_{r_c}(\mathbf s-\mathbf r)g_{r_c}(\mathbf s-\mathbf r')
=\exp[-|\mathbf r-\mathbf r'|^2/(4r_c^2)]\).
This reduces to Eq.~\eqref{eq:cslpoint} for
\(\mu_i(\mathbf r)=m\delta(\mathbf r-\mathbf R_i)\). For Gaussian branch mass
densities
\[
  \mu_i(\mathbf r)=
  \frac{m}{(2\pi\sigma^2)^{3/2}}
  \exp\!\left[-\frac{|\mathbf r-\mathbf R_i|^2}{2\sigma^2}\right],
\]
the convolution gives
\[
  \mathcal I_{ij}=
  m^2\left(1+\frac{\sigma^2}{r_c^2}\right)^{-3/2}
  \exp\!\left[-\frac{|\mathbf R_i-\mathbf R_j|^2}
  {4(r_c^2+\sigma^2)}\right].
\]
Taking \(|\mathbf R_1-\mathbf R_2|=D(t)\) gives
Eq.~\eqref{eq:cslfinitewidth}.
The isotropic three-dimensional estimate used in the main text is the
conservative choice for the present parameter table.

\subsection{Shot-noise criterion}
\label{app:shotnoise}

Each run ends with a binary molecular-state readout. Assign the two outcomes
the values \(z=+1\) and \(z=-1\). At the most sensitive Ramsey phase,
\(\langle z\rangle=0\) and \(\langle z^2\rangle=1\), so the sample mean
\(\bar z\) has rms uncertainty \(1/\sqrt N\). The fitted Ramsey fringe amplitude
is the normalized amplitude of this population difference, so its shot-noise
scale is
\[
  \delta V\simeq \frac{1}{\sqrt N}.
\]
A collapse exponent \(\Lambda_{\rm c}\) produces
\[
  \Delta V=\e^{-\Lambda_{\rm loss}}\left(1-\e^{-\Lambda_{\rm c}}\right).
\]
The exact \(5\sigma\) shot requirement is
\[
  N_{5\sigma}=
  \frac{25\e^{2\Lambda_{\rm loss}}}
  {\left(1-\e^{-\Lambda_{\rm c}}\right)^2}.
\]
For the apparatus-exposure benchmark,
\(\Lambda_{\rm c}=\Gamma_{\rm exp}J_{\rm exp}\).
In the small-exponent limit, the \(5\sigma\) condition gives
\[
  \e^{-\Lambda_{\rm loss}}\Gamma_{\rm exp}J_{\rm exp}
  \simeq \frac{5}{\sqrt{N}},
\]
and therefore Eq.~\eqref{eq:gammaexpmin}. For
\(\Lambda_{\rm c}=\lambda K_{\rm CSL}\ll1\), the same argument gives
Eq.~\eqref{eq:lambdamin}.

% !TEX root = quantum_reversible_measurement.tex

\end{document}